\documentclass[manuscript]{aastex}
\usepackage{emulateapj5}
\usepackage{epsf}
\usepackage{psfig}
\usepackage{graphicx}
\usepackage{lscape}
\usepackage{natbib}
\usepackage{enumerate}
\usepackage[flushleft]{threeparttable}

\shorttitle{}
\shortauthors{Rich et al.}
\let\MakeUppercase\relax

\begin{document}
\title{Thermal Infrared Imaging and Atmospheric Modeling of VHS J125601.92-125723.9 \MakeLowercase{b}: Evidence for Moderately Thick Clouds and Equilibrium Carbon Chemistry in a Hierarchical Triple System}

\author{Evan A. Rich\altaffilmark{1},
Thayne Currie\altaffilmark{2},
John P. Wisniewski\altaffilmark{1},
Jun Hashimoto\altaffilmark{3},
Timothy D. Brandt\altaffilmark{4,5},
Joseph C. Carson\altaffilmark{6},
Masayuki Kuzuhara\altaffilmark{7},
Taichi Uyama\altaffilmark{8}}

\altaffiltext{1}{Homer L. Dodge Department of Physics, University of Oklahoma, 440 W. Brooks St., Norman, OK 73071, USA}
\altaffiltext{2}{National Astronomical Observatory of Japan, Subaru Telescope}
\altaffiltext{3}{Astrobiology Center, National Institutes of Natural Sciences, National 
Astronomical Observatory of Japan, 2-21-1 Osawa, Mitaka, Tokyo 181-8588 Japan}
\altaffiltext{4}{Astrophysics Department, Institute for Advanced Study, Princeton, NJ, USA}
\altaffiltext{5}{NASA Sagan Fellow}
\altaffiltext{6}{Department of Physics and Astronomy, College of Charleston, 66 George Street., Charleston, SC 29424, USA}
\altaffiltext{7}{Department of Earth and Planetary Sciences, Tokyo Institute of Technology, 2-12-1 Ookayama, Meguro-ku,
Tokyo 152-8551, Japan}
\altaffiltext{8}{Department of Astronomy, The University of Tokyo, 7-3-1, Hongo, Bunkyo-ku, Tokyo, 113-0033, Japan}

\begin{abstract}

We present and analyze Subaru/IRCS $L^\prime$and $M^\prime$ images of the nearby M dwarf 
VHS J125601.92-125723.9 (VHS 1256), which was recently claimed to have 
a $\sim$11 M$_{J}$ companion (VHS 1256 b) at $\sim$102 au separation. 
Our adaptive optics images partially resolve the central star into a binary, whose components are nearly equal in brightness 
and separated by 0$\farcs$106 $\pm$ 0$\farcs$001. 
VHS 1256 b occupies nearly the same near-infrared color-magnitude diagram position as HR 8799 bcde and has a comparable $L^\prime$ brightness. 
However, it has a substantially redder $H$ - $M^\prime$ color, implying a relatively brighter $M^\prime$ flux density than for the HR 8799 planets and suggesting that non-equilibrium carbon chemistry may be less significant in VHS 1256 b. 
We successfully match the entire SED (optical through thermal infrared) for VHS 1256 b to atmospheric models assuming chemical equilibrium, models which failed to reproduce HR 8799 b at 5 $\mu m$. 
Our modeling favors slightly thick clouds in the companion's atmosphere, although perhaps not quite as thick as those favored recently for HR 8799 bcde. 
Combined with the non-detection of lithium in the primary, we estimate that the system is at least older than 200 Myr and the masses of the stars comprising the central binary are at least 58 $M_{J}$ each.  Moreover, we find some of the properties of VHS 1256 are 
inconsistent with the recent suggestion that it is a member of the AB Dor moving group.
Given the possible ranges in distance (12.7 $pc$ vs. 17.1 $pc$), the lower mass limit for VHS 1256 b ranges from
10.5 $M_J$ to 26.2 $M_J$.
Our detection limits rule out companions more massive than VHS 1256 b exterior to 6--8 au, placing significant limits on and providing some evidence against a second, more massive companion that may have scattered the wide-separation companion to its current location.
VHS 1256 is most likely a very low mass (VLM) hierarchical triple system, and could be the third such system in which all components reside in the brown dwarf mass regime.   

\end{abstract}
\section{Introduction}

Exoplanet surveys have recently measured the frequency of exoplanets as a function of a host of 
parameters like stellar mass, metallicity, orbital separation, and planetary mass \citep{win15}. 
These parameters help to inform our understanding of how and where exoplanets form.  The observed 
frequency of gas giants at small ($<$2 au) separations rises from $\sim$3\% for dM stars to $\sim$14\% 
for solar metallicity A-type stars; this dearth of massive planets around dM stars is consistent with 
theoretical predictions \citep{lau04} of the core accretion model \citep{pollack1996}.  At 
larger orbital separations (10s-100s of au), the frequency of gas 
giants around dM stars is $<$6\% \citep{bowler2015}.  Recent results from 
Kepler have shown both that the frequency of small mass planets at short orbital periods increases
around low mass stars \citep{bor11,how12} and there is a lack of planets larger than 2.5 R$_{\earth}$ 
surrounding dM stars at short orbital periods.

High contrast imaging investigations have similarly begun to discover gas giant exoplanets 
located at large orbital separations from their stars
 (Fomalhaut b, \citealt{kal08}, $\beta$ Pictoris b, \citealt{lagrange2009,lag10}; HR 8799 bcde, \citealt{mar10}; $\kappa$ And. b, \citealt{car13}; 51 Eri b ,\citealt{mac15}; HD 100546 bc ,\citealt{quanz13}, \citealt{cur15}).  
While most of these directly imaged gas giants surround early-type stars, detections have been reported around Solar-analogs (GJ 504b, \citealt{kuz13}) and dM stars (e.g. ROXs42B, \citealt{cur14}; GU Psc, \citealt{nau14}). 
Yet the formation mechanisms responsible for these systems is still under debate.
A growing number of objects with wide orbits and modest mass ratios (eg. HD 106906b, \citealt{bai14}; ROXs42Bb, \citealt{cur14}; 1RXSJ1609, \citealt{Laf2008}; 2M J044144, \citealt{trodorv2010})
have led to suggestions that the planetary companion formed via a binary star-like process rather than the core accretion process (\citealt{Low1976}, \citealt{bate2009}, \citealt{bra14}).

Although binary stars are common (e.g. \citealt{rag10}), our understanding of the frequency of exoplanets around binaries and higher order systems remains limited.  Since the discovery of the first exoplanet surrounding a binary host (Kepler-16b, \citealt{doy11}), less than a dozen similar systems have been discovered by Kepler (\citealt{win15} and references therein).  The analysis of publicly available Kepler data led \citet{arm14} to conclude that the frequency of planets with R $>$ 6R$_{\earth}$ on 
periods of less than 300 days was similar to that of single star rates; however, this conclusion is 
critically dependent on the assumed planetary inclination distribution.  While at least one bona fide planetary mass companion orbiting a binary has been imaged \citep[ROXs 42Bb][]{sim95,rat05,cur14} most dedicated direct imaging surveys for gas giant planets around binaries have not yielded any firm detections to-date \citep{tha14}.

Recently, \citet{gau15} reported the detection of a planetary mass (11.2 $^{+9.7}_{-1.8}$ M$_{J}$) companion at a projected separation of 102 $\pm$9 au from its host star VHS J125601.92-125723.9 (hereafter VHS 1256), described as a M7.5 object with an inferred mass from its bolometric luminosity of 73$^{+20}_{-15}$ M$_{J}$, placing it near the hydrogen burning limit.    The primary was estimated to have an age of 150-300 Myr from both kinematic membership in the Local Association and lithium abundance.  
At a distance of $12.7 \pm 1.0$ $pc$ measured from trigonometric parallax \citep{gau15}, this made VHS 1256 the closest directly imaged planetary mass system to the Earth.
\citet{Stone2016} reported a greater distance to VHS 1256 of $17.1 \pm 2.5$ $pc$ based on spectrophotometry of the system. 

From the standpoint of substellar atmospheres and atmospheric evolution, VHS 1256 b is a particularly unique object.  Its near-infrared properties resemble those of the HR 8799 planets and a select few other young (t $\lesssim$ 30 Myr) and very low mass (M $\lesssim$ 15 $M_{\rm J}$) substellar objects, occupying  roughly the same near-infrared color-magnitude space \citep{gau15,Faherty2016}: a continuation of the L dwarf sequence to fainter magnitudes and cooler temperatures.   Indeed, as shown by atmosphere modeling, the near-infrared properties of objects like HR 8799 bcde and 2M 1207 B reveal evidence for thicker clouds than field brown dwarfs of the same effective temperatures \citep{Currie2011}.  VHS 1256 b then offers a probe of clouds at ages intermediate between these benchmark objects and Gyr-old field objects and thus some insights into the atmospheric evolution of low-mass substellar objects.

Furthermore, non-equilbrium carbon chemistry 
can be probed by new thermal infrared photometry, in particular at $M^\prime$ \citep[e.g.][]{Galicher2011}\footnote{The existing W2 photometry reported in \citet{gau15} covers a far wider bandpass (4--5 $\mu m$).  Much of this wavelength range is far less sensitive to carbon monoxide opacity at relevant temperatures that is a tracer of non-equilbrium carbon chemistry, while $M^\prime$ is far more (uniquely) sensitive (e.g. see Figure 7 in \citealt{cur14B}).}.  New thermal infrared data for VHS 1256 b allows us to assess the evidence for non-equilibrium chemistry for the objects at/near the deuterium burning limit and at ages older than HR 8799 bcde.
 
In this work, we present new adaptive optics imagery of VHS 1256, providing the first detections of its wide-separation companion in major thermal IR broadband filters, $L^\prime$ and $M^\prime$.  We use these mid-infrared photometric points and optical and near-infrared photometry from \citet{gau15} to perform the first atmospheric (forward) modeling of VHS 1256 b and the first assessment of how its thermal IR properties (e.g. carbon chemistry) compare to younger planet-mass objects with similar near-IR colors.
Additionally, we report our independent determination of the primary's binarity, also reported in \citet{Stone2016}, following our original work \citep{rich2015} with additional analyses. 
We will adopt the same nomenclature as Stone et al. \citeyear{Stone2016}, referring to the close partially resolved binary as VHS 1256 A and B, and the wide companion as VHS 1256 b.

After discussing our observations in Section \ref{sec:obs}, we search for new companions around VHS 1256 and investigate the binarity of VHS 1256 in Section \ref{sec:point}.
Next, we discuss improved photometry of VHS 1256 A, B, and b at $L^\prime$and $M^\prime$ in Section \ref{sec:photometry}. Using the new L$^\prime$ and M$^\prime$ photometry, we assess the atmospheric properties of VHS 1256 b in Section \ref{sec:atmo}.
Finally, we discuss the implications of our study in Section \ref{sec:Discussion}.

\section{Observations and Reductions} \label{sec:obs}

We observed VHS 1256 in $L^\prime$($\lambda_{center}$ = 3.77 $\mu$m) and $M^\prime$ ($\lambda_{center}$ = 4.68 $\mu$m) bands on 2015 June 6 using the Infrared Camera and Spectrograph (IRCS; \citet{kob00}) on the 8.2m Subaru Telescope, along with the Subaru AO-188 adaptive optics (AO) system \citep{hay08,hay10} in laser guide star (LGS) mode.  All observations were made with 
IRCS having a plate scale of 20.57 mas pix$^{-1}$.  For the L' band, we utilized a 0.1 second exposure time along with 
100 co-adds to achieve 10 second exposures per dither frame.  We obtained a total of 13 dither sets, using 
a 5 point dither per set, yielding a total on-source integration time of 650 seconds.  For our $M^\prime$-band 
imaging, we utilized a 0.06 second exposure time along with 100 co-adds to achieve 6 second exposures 
per dither frame.  We observed VHS 1256 for a total of 3 dither sets, using a 5 point dither per set, yielding a total on-source integration time of 90 seconds.  All data were obtained in angular differential imaging (ADI) mode; a total field rotation of 17$^{\circ}$.4 and 2$^{\circ}$.8 was achieved in $L^\prime$and $M^\prime$ respectively. 

We utilized two approaches to detect VHS 1256 b.  First,
we simply derotated each image to true north after sky 
subtraction.  We inspected the quality of each image set, 
and determined that the AO performance was subpar during 7 $L^\prime$frames (70 seconds of integration); hence, we 
removed these data before median combining frames.
Using an identical process, we reduced the $M^\prime$ data and removed 6 of the 15 frames (54 seconds of integration) due to subpar AO performance.

Second, we utilized advanced PSF subtraction methods and a different approach with image combination to compensate for the shorter integration time and brighter sky background in $M^\prime$ and obtained slightly deeper limits in $L^\prime$.  We use A-LOCI as in \citet{Currie2012,cur15b} with a large optimization area (500 PSF footprints) and rotation gap (1.5 times the diffraction limit at L$^\prime$ and M$^{\prime}$) combined with a moving pixel mask to flatten the background without removing signal from VHS 1256 b\footnote{as shown by \citet{Galicher2011}, advanced methods usually used for point-spread function (PSF) subtraction can in some cases better flatten the sky background, especially in M$^\prime$.  
As VHS 1256 b is located about 8\arcsec{} from the primary, even a small 2$^{\circ}$.8 parallactic motion is sufficient to apply PSF subtraction methods like LOCI \citep{Lafreniere2007a}}.
We then applied a 4 $\lambda$/D spatial filter and combined the derotated data to obtain a mean image using a 3-$\sigma$ outlier rejection. 
This allowed us to obtain a higher signal-to-noise detection of VHS 1256 b, which was barely detectable in $M^\prime$ at the 3-$\sigma$ level using the first image processing method, obtaining about a 40\% deeper background limit at $L^\prime$.

We utilized observations of the $L^{\prime}$ standard FS 138 \citep{van96} obtained on 2015 June 7 using NGS and 
observations of the $M^{\prime}$ standard HR 5384 \citep{van96} obtained on 2015 June 8 without AO to transform our 
photometry to a standard system.  We did not obtain successful images of a dense stellar cluster that would be needed to construct a robust distortion correction for our 
$L^\prime$and $M^\prime$ imagery; hence, we have included no such correction 
to our data.  Our lack of a distortion correction introduces some uncertainty into VHS 1256 b's 
separation, but does not affect our other results.

\section{Search for Additional Companions}\label{sec:point}
\subsection{Point Source Detections and Detection Limits in the $L^\prime$and $M^\prime$ Field of View}
\begin{figure}[hb]
\epsscale{0.9}
\plotone{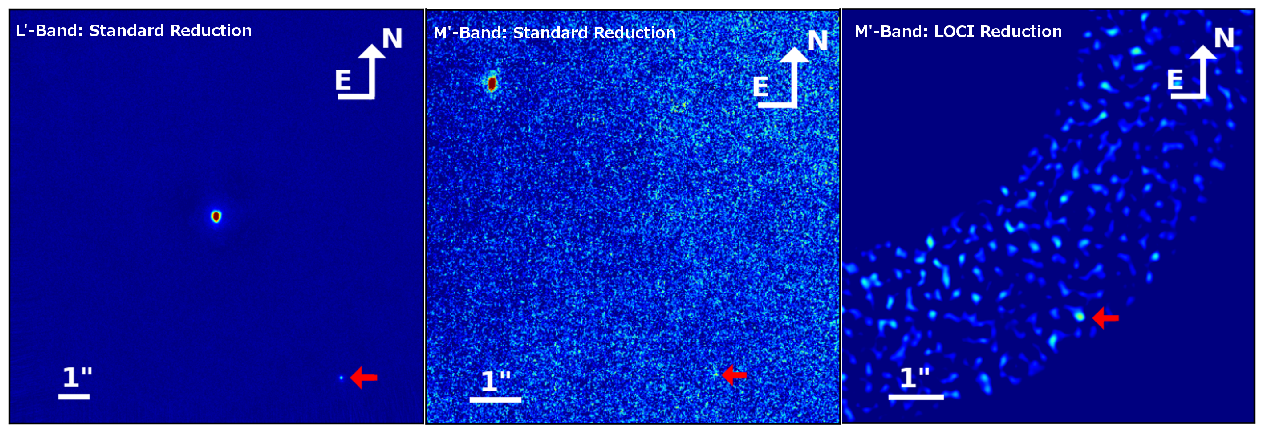}
\caption{Subaru/IRCS $L^\prime$-band ($\sim$16$\farcs$4 x $\sim$16$\farcs$4 FOV; left panel), $M^\prime$-band ($\sim$9$\farcs$5 x $\sim$9$\farcs$5 FOV; middle panel), and $M^\prime$-band ($\sim$8$\farcs$5 x $\sim$8$\farcs$4 FOV; LOCI reduction; 3 sigma filter applied; right panel) AO 
observations of the VHS 1256 system. Left and center panels show the full usable FOV with the partially resolved binary in the left and center panel and VHS 1256 b previously detected by \citet{gau15} in all three panels. The red arrow depicts the location of VHS 1256 b. Note that no other point sources are detected in the FOV. The data are plotted on a linear intensity scale.}
\label{fig:L_M}
\end{figure}

Our fully reduced $L^\prime$- and $M^\prime$-band imagery (Figure \ref{fig:L_M}) clearly reveals 
the presence of both the composite source of VHS 1256 A and B (A+B) and VHS 1256 b reported by \citet{gau15}.   We detect the companion at SNR $\sim$ 130 (100) in the $L^\prime$ data  and at 4.5 (3) in the $M^\prime$ data using the A-LOCI (classical) reduction, where the latter detection is roughly comparable to the detection significance of HR 8799 bcd in \citet{Galicher2011}.  We do not identify any additional point sources.

We determined the centroid positions of both objects, and found that VHS 1256 b is located 8$\farcs$13 $\pm$ 0$\farcs$04 from VHS 1256 A+B at a position angle of $217.8^{\circ} \pm 0.3^{\circ}$ in $L^\prime$and 8$\farcs$17 $\pm$ 0$\farcs$04 at a position angle of $217.8^{\circ} \pm 0.3^{\circ}$ in $M^\prime$ (Table \ref{tbl:properties}).
This separation is consistent with the angular separation observed by \citet{gau15} of 8$\farcs$06 $\pm$ 0$\farcs$03 at a position angle of $218.1^{\circ} \pm 0.2^{\circ}$. Note that we did not utilize a distortion correction. 

To compute the 5-$\sigma$ point source detection limit in $L^\prime$, we followed standard methods used in high-contrast imaging data \citep[e.g.][]{cur15b}. We convolved the image with a gaussian profile having a FWHM 
set by the observed FWHM of VHS 1256 b and determination the robust standard deviation of convolved pixels at each angular separation.
The limiting 5-$\sigma$ point source detection limits 
of our $L^\prime$ imagery based on this method is roughly 16.4 mag exterior to 2\farcs{}5 (r$_{\rm proj}$ $\sim$ 32 au for $d$ = 12.7 $pc$), degrading to VHS 1256 b's brightness at 0\farcs{}5.
Due to the thermal background, limited integration time, and the partial FOV coverage, the $M^\prime$ imagery do not provide stringent constraints on the presence of additional point sources compared to the $L^\prime$imagery.

\subsection{Analysis of the Primary: Evidence of Multiplicity}

\begin{figure}[hb]
\epsscale{0.9}
\plotone{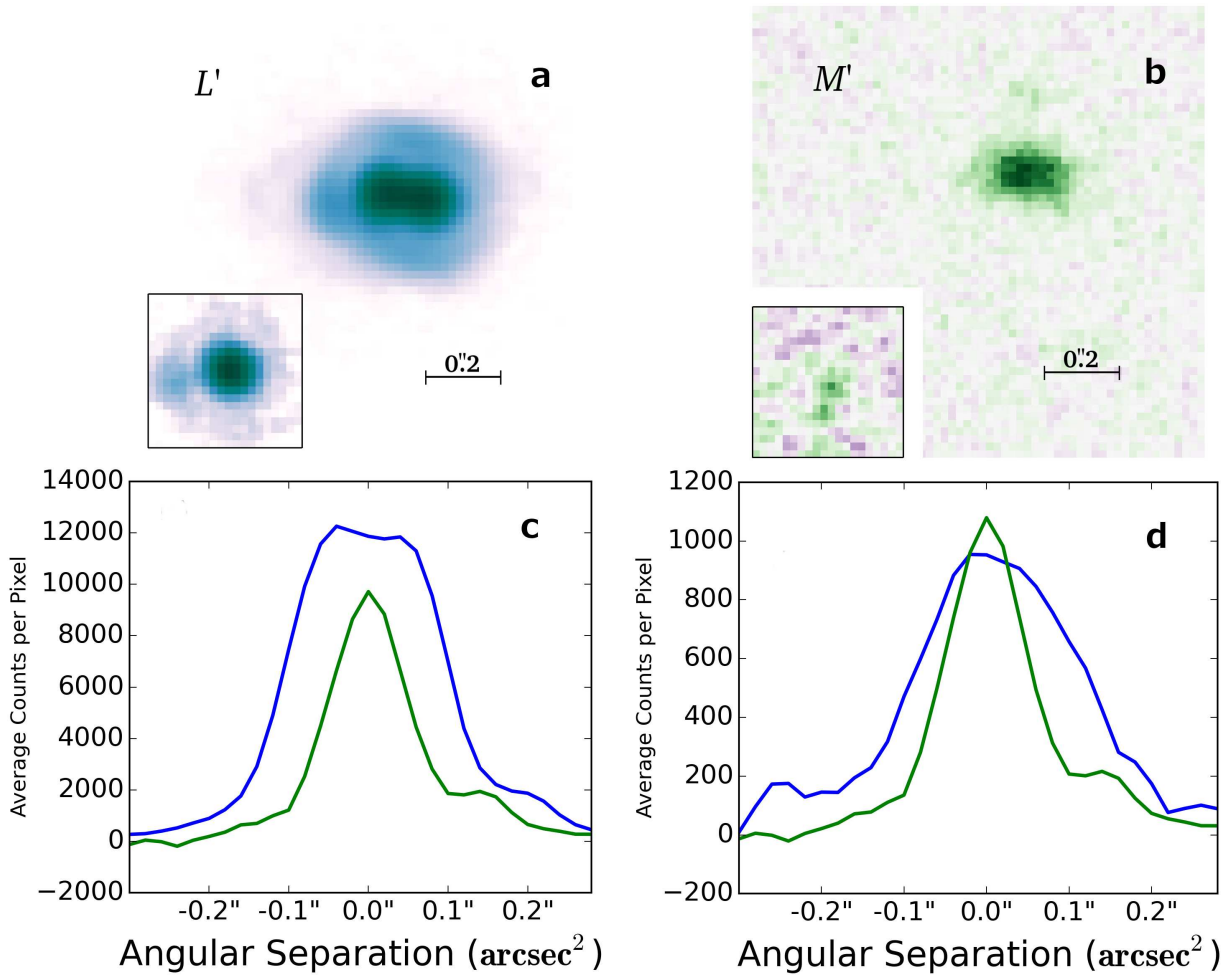}
\caption{The ellipticity of VHS 1256 A and B in $L^\prime$(panels a and c) and $M^\prime$ (panels b and d) can be seen above.  
Panels a and b show a FOV (1$\farcs$3 X 1$\farcs$3) of VHS 1256 A and B in their respective band-passes with north to the left and east up. The insert in panels a) and b) is of VHS 1256 b used as the PSF. Note that the panel b) insert used the A-LOCI reduction rather than the first reduction method. Panels c and d are crosscuts along the major axis of the central source (blue) and PSF (green), averaged over 5 pixels.
The clear double-peak profile in panel c illustrates that we partially resolve the central source in $L^\prime$.}
\label{fig:contour}
\end{figure}

It is apparent from visual inspection of Figure \ref{fig:L_M} that the VHS 1256 A+B is much more 
elongated in both $L^\prime$and $M^\prime$ -bands than VHS 1256 b.  In fact, simple gaussian fits to the 
data reveal that the FWHM of the central star is significantly broader than the companion in both 
band-passes.  
Zooming in on the region around VHS 1256 A+B, Figure \ref{fig:contour} demonstrates that the central star is comprised of 
two marginally resolved sources.  Using a modified elliptical least squares minimization fitting
routine written by Nicky van Foreest \footnote{http://nicky.vanforeest.com/misc/fitEllipse/fitEllipse.html}, we compute the average 
ellipticity (e = ((a$^{2}$ - b$^{2}$)/a$^{2}$)$^{0.5}$)) of our sources.  The average ellipticity 
of the central star, $0.85 \pm 0.02$ and $0.825 \pm 0.005$ in the $L^\prime$and $M^\prime$ bands 
respectively, is significantly greater than the average ellipticity observed for the companion 
($0.48 \pm 0.03$ in $L^\prime$).  These results all indicate that the central star is clearly comprised 
of the superposition of two separate sources.  

To help ascertain the probability that the observed multiplicity of VHS 1256 A and B arises from 
chance alignment of VHS 1256 with a background source, we utilize archival WISE W1-band imagery, 
which has a similar band-pass as our $L^\prime$-band imagery.  The local density of W1 sources 
between 10.2 and 10.8 mag is $\sim$ 34 $deg^{-2}.$
Hence, the chance alignment of two sources, within 0$\farcs$106 is $5\times10^{-6} \%$.

\section{Photometry}\label{sec:photometry}
Aperture photometry was performed on all components of the VHS 1256 system. Both the 
L$^{\prime}$ photometry of VHS 1256 A+B (9.757 $\pm$ 0.04 magnitude) and VHS 1256 b (12.93 $\pm$ 0.02 magnitude) and 
the M$^{\prime}$ photometry of VHS 1256 A+B (9.65 $\pm$0.04 magnitude) and VHS 1256 b (12.66 $\pm$0.26 magnitude) 
are broadly consistent with the literature \citep{gau15} WISE W1 and W2 photometry (Table \ref{tbl:photometry}).
We utilized a 30 pixels radius for aperture photometry. In addition to the standard uncertainty terms (poisson noise and standard deviation of background), we included read noise, shot noise, and the errors from the zero point terms.  We do not include error terms due to variations in the atmospheric transmission during 
our observations, or due to variations that arise between the use of NGS versus LGS observing modes.  Note that our $L^\prime$and $M^\prime$ VHS 1256 b photometry improves on the respective WISE W1 and W2 errors previously presented.

The relative astrometry and flux of VHS 1256 A and B were constrained by fitting a linear combination of two PSFs to our $L'$ images; we adopt VHS 1256 b as the PSF.
We first align the dithered images of the central binary by maximizing their cross-correlation and then combine them by averaging.  
Averaging the frames is optimal in the case of Gaussian errors; in our case, it reduces the noise by nearly a factor of 2 compared with a median combination.

We use bicubic spline interpolation to translate the PSF of VHS 1256 b.  We then simultaneously adjust the positions and normalizations of two PSFs to minimize
\begin{equation}
\chi^2 = \sum \frac{\left( {\rm PSF}_1 + {\rm PSF}_2 - {\rm data}\right)^2}{\sigma^2_{\rm ph} + \sigma^2_{\rm bg}}
\label{eq:chi2}
\end{equation}
where $\sigma^{2}_{\rm ph}$ is the variance from photon 
noise and $\sigma^2_{\rm bg}$ is the 
variance from read noise and the thermal background.  We estimate $\sigma^2_{\rm ph}$ from 
the instrument gain and measure $\sigma^2_{\rm bg}$ from empty regions of the image.  The 
variance is the sum of variance in the image and scaled variance from the noisy companion 
PSF.  In practice, most of the noise arises from scaling the companion PSF, as the 
noise increases linearly with the scaling factors.

A good fit should have $\chi^2 \approx N_{\rm pix}$, where $N_{\rm pix}$ is the number 
of pixels in the region being fit.  With a simple estimate of the background and
photon noise, we find that our best-fit $\chi^2$ is about four times the 
number of pixels; we therefore increase our estimated uncertainties to 
achieve a reduced $\chi^2$ of unity.  After rescaling our noise to achieve 
a minimum $\chi^2 = N_{\rm pix}$, we compute our confidence regions by 
integrating the likelihood 

\begin{equation}
{\cal L} = \exp \left[ -\chi^2/2 \right]
\label{eq:like}
\end{equation}

and computing the regions containing 68\% of the likelihood (for our 1$\sigma$ confidence intervals).  

\begin{figure}[ht]
\epsscale{0.9}
\plotone{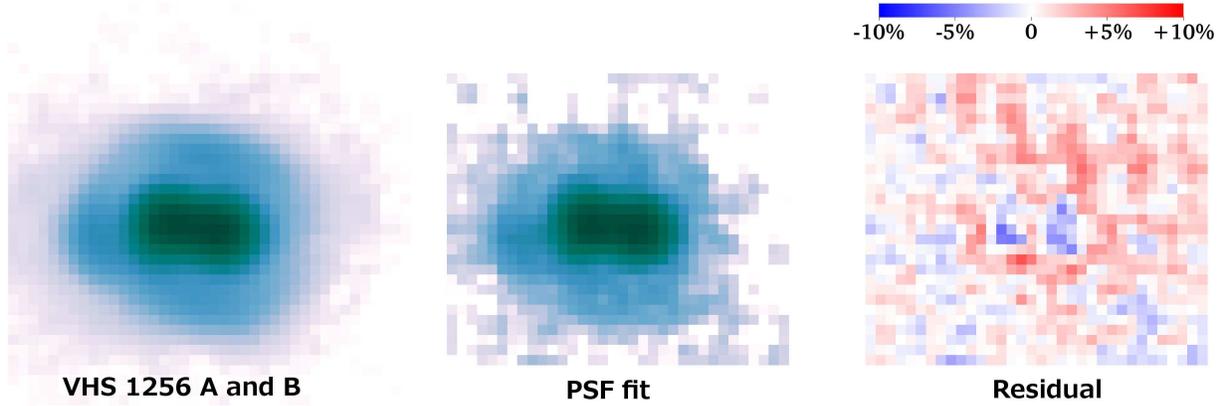}
\caption{The above shows PSF fitting of VHS 1256 A and B utilizing our $L^\prime$-band data. The left panel shows a view of VHS 1256 A and B. The middle panel shows the best fit result using VHS 1256 b as an observed PSF. The right panel shows the residual of the PSF fit.}
\label{fig:vhs}
\end{figure}

Figure \ref{fig:vhs} shows the results of this fit: the observed $L'$ intensity of VHS 1256 A and B (left panel), the best-fit translated and scaled linear combination of companion 
PSFs (middle panel), and the residuals (right panel), expressed in units of the peak intensity of the original data.  The fit is good to a few percent near the PSF cores, comparable to the expected uncertainties from the noise and from interpolations.  Spatial correlations 
in the noise are visible in the residual image even far from the PSF core; it is these correlations that force us to scale our errors.  A full treatment of the problem 
would modify $\chi^2$ 
(Equation \ref{eq:chi2}) to account for the data's non-diagonal covariance matrix.
Table \ref{tbl:properties} lists our fitted parameters: a separation 
of $0.\!\!''106 \pm 0.\!\!''001$ ($5.18 \pm 0.05$ pixels), a position angle of $-7.\!\!^\circ6 \pm 0.\!\!^\circ5$, and a flux ratio of $1.03 \pm 0.01$.  We thus obtain 1$\%$ measurements of the relative photometry and separation.  

\section{Atmospheric Properties of VHS 1256 \MakeLowercase{b}}
\label{sec:atmo}
The discovery work by \citet{gau15} suggested that VHS 1256 b, based on its spectral shape and color-magnitude diagram position, is a red L/T transition object similar to HR 8799 bcde \citep[e.g.][]{Currie2011,Bonnefoy2016}.  
With SED information spanning the red optical to mid-infrared, we can more thoroughly compare VHS 1256 b's properties to those of other substellar objects and fit atmosphere models with a range of assumptions about clouds to reproduce VHS1256 b's spectrum.  Many free floating L/T dwarfs with red optical to near-infrared photometry and spectra have been studied \citep[e.g.][]{knapp2004}.   However, even considering young brown dwarfs as well as the field, VHS 1256 b occupies an extremely sparsely populated near-infrared color-magnitude diagram position shared by planet mass objects like HR 8799 bcde and 2M 1207 B \citep{Faherty2016}.   
Furthermore, these objects are young, typically less than $\sim$ 30 $Myr$ old \citep{Faherty2016}, while VHS 1256 is at least older than 200 $Myr$ \citep[][sect 6.2 of this work]{Stone2016}.

While the similarity of VHS 1256 b's near-infrared colors to those of HR 8799 bcde suggest that their atmospheric properties likewise share some similarity (e.g. perhaps thick clouds), the addition of thermal infrared data in this work allows us to further quantify this feature and assess whether or not VHS 1256 b shows clear evidence for non-equilibrium carbon chemistry.   Compared to HR 8799 bcde, VHS 1256 b's available suite of photometry extends far bluer, into the optical.  Thus, the companion provides a new test of atmospheric models used to reproduce at least some bona fide directly-imaged planets.    Furthermore, as shown in \citet{Faherty2016}, young L dwarfs tend to follow a reddened version of the field sequence.   Comparing VHS 1256 A+B's colors to those of the field sequence and young associations like AB Dor can then provide a very coarse assessment of the system's youth.
  
  \subsection{Mid-Infrared Colors of VHS 1256 A, B, and b}
\begin{figure}[ht]
%\centering
\includegraphics[scale = 0.35]{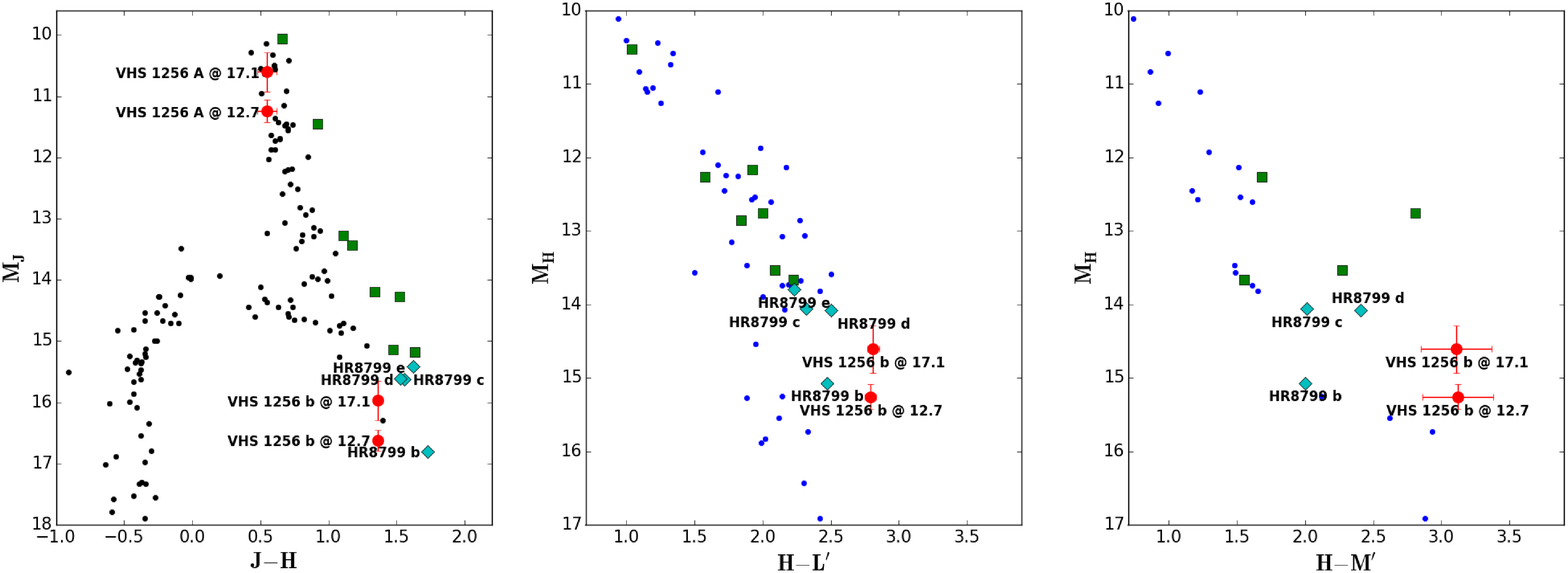}
\caption{Near-IR color magnitude diagrams in J-H (right panel) and H-$M^\prime$ (left panel) space are shown. For both panels, the red circles are VHS 1256 A and b for both 12.7 and 17.1 $pc$ distances (\citep{gau15,Stone2016}), the blue circles are HR 8799 bcde and L/T dwarfs with $M^\prime$ photometry \citep{Legget2002,cur14B}, the green circles are members of the AB Dor moving group \citep{Faherty2016}, and 
the black cicles are M/L/T field dwarfs \citet{dupuy2012}. Note that VHS 1256 b is consistent in color with HR 8799 b in J-H space (right panel) but VHS 1256 b has a much redder color in H-$M^\prime$ space (left panel).}
\label{fig:cmd}
\end{figure}

  Figure  \ref{fig:cmd} compares VHS 1256 A, B, and b's near-infrared and mid-infrared color magnitude diagrams (CMDs) to the field sequence, planetary companions like HR 8799 bcde, and substellar objects in AB Dor ($t$ $\sim$ 125 $Myr$).  The primary components to VHS 1256 appear indistinguishable from the field sequence and bluer than AB Dor members.  Even if VHS 1256 b is at 17.1 $pc$ and thus intrinsically brighter, it still appears somewhat ``under luminous", along an extension of the L dwarf sequence to fainter magnitudes and presumably lower effective temperatures.   Depending on the distance, VHS 1256 b's $L^\prime$/H-$L^\prime$ position is either consistent with HR 8799 b's or appears more comparable to the inner three planets.  
  
In contrast, VHS 1256 b is significantly brighter at $M^\prime$, relative to HR 8799 bcde.
VHS 1256 b's H-$M^\prime$ color is  0.6--1.1 magnitudes redder that HR 8799 bcd, suggesting that at $M^\prime$ it is roughly 1.7--2.7 times as bright.
The \citet{Leggett2010} and \citet{dupuy2012} compilations of M/L/T objects (Figure \ref{fig:cmd}) are more sparsely populated in $M^\prime$ for late L and T dwarfs making comparisons with the field sequence difficult. 
As the faint $M^\prime$ brightnesses for L/T transition objects and the HR 8799 planets signaled evidence for non-equilibrium carbon chemistry \citep{Galicher2011}, VHS 1256 b's brighter $M^\prime$ flux density may indicate that non-equilibrium carbon chemistry is not significant for every L/T transition object.

  \subsection{Atmospheric Modeling: Methodology}
We perform atmospheric forward modeling to estimate VHS 1256 b's temperature, surface gravity, and radius, and to see if its photometry can be reproduced by 
atmospheric models in chemical equilibrium.
We follow the studies of \citet{Currie2011} and \citet{Madhusudhan2011} for HR 8799's planets and \citet{Burrows2006} for field brown dwarfs, comparing the data with atmosphere models from A.
Burrows covering a range of parameterized cloud prescriptions but all assuming chemical equilibrium.  

We consider the ``E60" models, which, have a modal particle size of 60 $\mu m$ and clouds sharply truncated at depth, well below the planet photosphere.  We then consider ``A60" and ``AE60" cloud models, which have an identical modal particle size but simulate ``very thick" and "thick" cloud model presciptions (see \citealt{Burrows2006} and \citet{Madhusudhan2011} for details).  
From the model fitting, we identify those consistent with the data to within 3-$\sigma$ using a simple $\chi^{2}$ threshold, setting the minimum photometry uncertainty to be 10\% as in \citet{Currie2011}.  
This method then yields a plausible range of temperature, surface gravity, radius, luminosity and mass from the ensemble of acceptably-fitting models. 

We perform two sets of model fits, one where we allow the planet radius to freely vary and another when we pin it to values implied in the \citet{Burrows2001} evolutionary models.  
We assess how the system's uncertain distance affects our best-fit derived by performing fits, for the fixed-radius case, assuming either distances (12.7 and 17.1 $pc$). 

\subsection{Atmospheric Modeling: Results}
\begin{figure}[ht]
%\centering
\includegraphics[scale = 0.34]{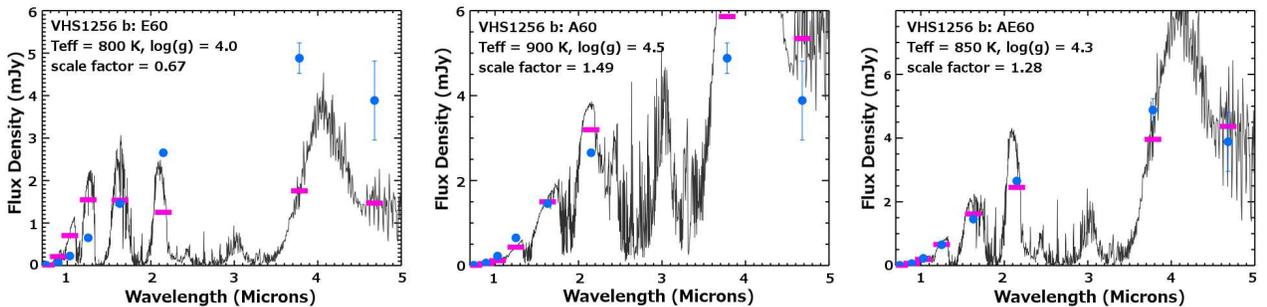}
\caption{Best-fit atmosphere models for the thin-cloud E60 model (left panel), the very-thick cloud A60 model (middle panel), and the thick cloud AE60 model (right panel).  The (very) thick cloud models accurately reproduce the optical to mid-IR SED of VHS1256 b. Black lines is the model spectra, the magenta points are the model spectra applied to the appropriate filter, and the blue points are photometry that match the magenta filter points.}
\label{fig:sedmod}
\end{figure}

Figure \ref{fig:sedmod} displays best-fit models using each cloud model assumption and Tables \ref{tbl:atmosfit} and \ref{tbl:atmosfitfixed} summarize our model fitting results. VHS 1256 b's SED cannot be reproduced by the thin-cloud, E60 models: like HR 8799's planets, such models badly under-predict the companion's brightness at the shortest wavelengths and over-predict it in the thermal infrared \citep[see][]{Currie2011}.  The A60, very-thick cloud models yield a drastically improved fit at temperatures between 900 $K$ and 1200 $K$, where a model with $T_{eff}$ = 1000 K, log(g) = 4.5 is marginally consistent with the optical to mid-infrared SED at the 3-$\sigma$ confidence limit.

Models with slightly thinner clouds, AE60, fare better, yielding a wider range of temperature and surface gravity phase space able to match the data.  The best-fit values systematically skew towards lower temperatures and surface gravities, where the best fit model has $T_{eff}$ = 800 $K$ and log(g) = 3.8 when the radius can freely vary and has a slightly higher gravity (log(g) = 4.1) when the radius is fixed.

Assuming a distance of 17.1 $pc$, the implied luminosity of VHS 1256 b ranges between log(L/L$_{\odot}$) = $-4.79$ to $-4.95$. Adopting the 12.7 $pc$ distance yields a luminosity consistent with that previously estimated by \citet{gau15}: log(L/L$_{\odot}$) = $-5.06$ to $-5.24$. We will estimate the mass of VHS 1256 b by using these luminosities in section \ref{sec:Masses}. While successful at reproducing VHS 1256 b's SED, the derived model fit parameters could be revised by future modeling efforts.  For instance, the implied radii for best-fit models are generally larger than predicted for substellar object's with ages greater than 150 $Myr$ \citep[see][]{Baraffe2003}, and thus similarly the temperatures could be slightly larger than implied by our analysis.

\section{Discussion}
\label{sec:Discussion}
\subsection{Binarity of the Central Source}
Subaru/IRCS AO $L^\prime$and $M^\prime$ imagery has clearly revealed that the central source of 
the VHS 1256 system is comprised of two objects (Figure \ref{fig:L_M}) that have similar 
relative brightness ($L^\prime$= 10.5 and 10.54 magnitude respectively).  Such binarity is observed in 22$^{+6}_{-4}$\% of very low mass stars \citep{duc13}.  \citet{gau15} assigned 
the central source a spectral classification of M7.5, based on optical (M7.0) and IR (M8.0) spectral 
classifications.
We speculate that the minor differences in the optical versus IR spectral classifications derived by \citet{gau15} could be caused by minor 
differences in the spectral classifications of the binary components.  At the observed 
distance to VHS 1256 
($12.7 \pm 1.0$ $pc$; \citealt{gau15}), the $0\farcs103 \pm 0\farcs001$ projected separation 
between the binary components corresponds to a projected physical separation of 
$\sim$1.3 au.  Our results on the binarity of the central source are consistent with those independently and recently reported by \citet{Stone2016}.

\subsection{System Age and Component Masses}
\label{sec:Masses}
\begin{figure}[ht]
\includegraphics[scale = 0.4]{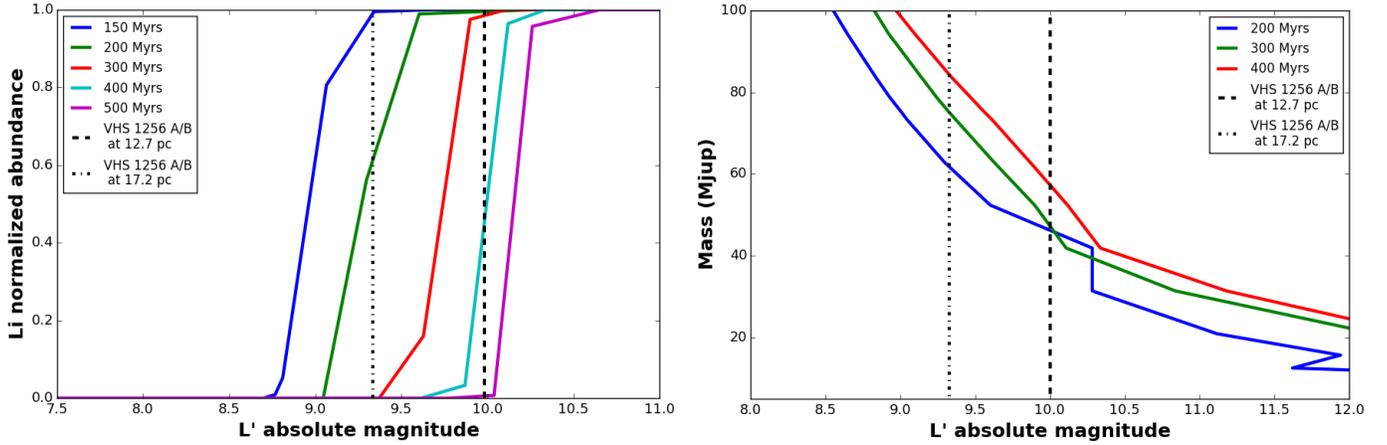}
%\centering
\caption{The above figures demonstrate the age of the VHS 1256 system (left panel), and the estimated lower mass limit of VHS 1256 A and B (right panel).  The left panel plots the normalized lithium abundance against the absolute L' magnitude taken from BT-Settl models \citep{all14} for a variety of ages.
The absolute magnitudes of VHS 1256 A and B (both 12.7 and 17.1 $pc$ distances) are taken from the PSF fit shown in table \ref{tbl:absolute}, plotted as the vertical dashed lines. Unresolved spectroscopy of stars A and B from \citet{gau15} showed no detection of lithium down to 30 $m\AA$, 
thus these stars must be old enough to have destroyed the initial lithium. 
The best lower limit ages based no the non-detection of lithium are $>$ 200 Myrs (17.1 $pc$) or $>$ 400 Myrs (12.7 $pc$). 
The right panel plots the sub-stellar mass ($M_J$) against the absolute L' magnitude taken from BT-Settl models. 
The solid lines are three different BT-Settl models: 200 Myrs for the 17.1 $pc$ distance, 400 Myrs for the 12.7 $pc$ distance, and 300 Myrs for the Gauza et al. (2015) upper age limit.
The estimated lower mass limit for VHD 1256 A and B for both 12.7 and 17.1 $pc$ distances is $>$ 58 $M_J$.}
\label{fig:lithium}
\end{figure}

\citet{gau15} suggested a system age of 150-300 Myr, based on the lack of observed Li in the system 
and kinematic age constraints from being a Local Association member.  
However, with the discovery that the central source is a binary \citep{Stone2016} and our independent verification of VHS 1256 A and B in the $L^\prime$-band, we can reassess the age limits of the system.
Using the nominal distance (12.7 pc) and the absolute magnitude (M$_{L'}$; $10.0 \pm 0.2$), the 
300 Myr upper limit age suggested by \citet{gau15} results in an inferred mass for VHS 1256 A or B of 47 M$_{J}$.
However such a mass would be too small to destroy Li \citep{all14} and produce the non-detection of this 
line (Figure \ref{fig:lithium}).  Rather, at this adopted distance the lower limit age of VHS 1256 must 
be $>$ 400 Myr to produce 
VHS 1256 A and B with our observed M$_{L'}$ and the lack of Li in the system's spectra.  If one assumes the 
new distance of 17.1 pc proposed by \citet{Stone2016} and the corresponding absolute magnitude 
of the central components (M$_{L'}$; $\sim9.4 \pm 0.3$), the lower age limit is $>$ 200 Myrs (Figure 5).
This is broadly consistent with the lower age limit proposed by \citet{Stone2016} of $280^{+40}_{-50}$. 
Note we used models from \citet{all14}, while Stone et al. used models from \citet{Chabrier2000}.

Stone et al. (\citeyear{Stone2016}) suggested that VHS 1256 was consistent with being a member of the 
AB Dor moving group, based on analysis of its UVW kinematics and a 66.85\% membership probability predicted by the 
BANYAN II software tool \citep{mal13,gag14}.
Our own investigation suggests that it still has a 28\% chance of being in the ``young field" (age up to 1 $Gyr$). Additionally, VHS 1256 b is a clear outlier in UWV space ($\sim$ 8 $\pm$ 1.7 km $s^{-1}$ from the core of AB Dor. (J. Gagne, pvt. comm.).
Furthermore, membership in the 149$^{+51}_{-19}$ Myr AB Dor moving group \citep{bel15} is inconsistent with the lower age limit of $280^{+40}_{-50}$ proposed by \citet{Stone2016} and marginally inconsistent with our lower limit of 200--400 $Myr$.  Moreover, the near-to-mid infrared colors of VHS 1256 A appear indistinguishable from those in the field and potentially bluer than AB Dor members (Figure \ref{fig:cmd}. Thus, it is not clear that 
VHS 1256 is a member of the AB Dor moving group, as suggested by \citet{Stone2016}.

As shown in Figure \ref{fig:lithium}, the minimum mass of each central component of VHS 1256 (A and B) 
is $>$ 58 M$_{J}$ for both of the distances discussed above.  This implies that the wide companion, VHS 1256 b, has a minimum mass ranging from 10.5 to 26.2 M$_{J}$, as shown in Table \ref{tbl:absolute}.
The large range is due to the uncertainty in the distance (12.7 or 17.1 pc) and the range in bolometric luminosities from atmospheric fitting (section \ref{sec:atmo}).
Though the lower estimate does dip below the deuterium burning limit, the companion is most likely in the brown dwarf regime.  

\subsection{Additional Companions and Formation}
\label{sec:Morecompanions}
\begin{figure}[ht]
%\centering
\includegraphics[scale = 0.7]{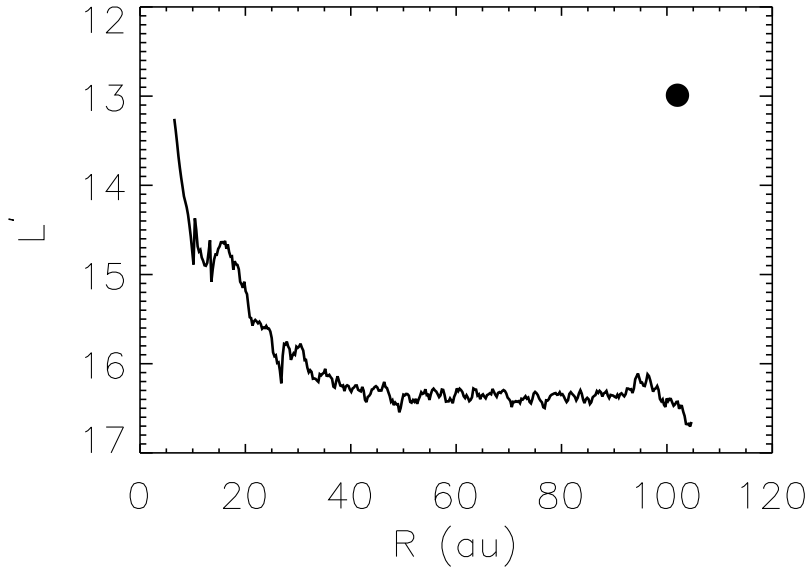}
\caption{Limiting background limit as a function of projected radius from VHS 1256 (A+B) for $L^\prime$ imagery with the A-loci reduction. The projected radius assumes a VHS 1256 distance of 12.7 $pc$ The black line is the 5-sigma background limit and the black dot is VHS 1256 b.}
\label{fig:lims}
\end{figure}

We detected no other point source companions in our field of view (FOV), $\sim$16$\farcs$5 x $\sim$16$\farcs$5 in $L^\prime$and $\sim$9$\farcs$3 x $\sim$9$\farcs$3 in $M^\prime$, down 
to our 5-$\sigma$ sensitivity limits shown in Figure \ref{fig:lims}.    
 of 13.2 ($L^\prime$; 12.5 mag at 17.1 pc).  
 Assuming a distance of 12.7 (17.1) $pc$, minimum system age of $>$ 400 (200) Myr, and no flux reversal at $L^\prime$ (i.e. that more massive objects are fainter), 
we can therefore exclude the presence of additional companions more massive than VHS 1256 b beyond 6 (8) au.   For most of the semi major axis space we probe, comparisons with \citet{Baraffe2003} imply that companions down to 3--5 $M_{\rm J}$ can be excluded if the system is 200--400 $Myr$ old.

Because we have failed to identify other substellar companions orbiting the primary, this severely restricts the possibility that VHS 1256 b was scattered to its present orbit by dynamical interactions with another, unseen planet.  Thus far, searches for close-in substellar companions to stars with imaged (near) planet-mass companions at 100--500 au have failed to identify potential scatterers, suggesting that this class of objects formed in situ either from protostellar disk or molecular cloud fragmentation \citep{Bryan2016}.  

Furthremore, the mass ratio (q) of VHS 1256 b (M$\sim$18.4 M$_{J}$; median lower limit between 10.2-26.2 M$_{J}$) to VHS 1256 A+B (M$\geq$116 $M_J$) is $\sim$0.16.
This mass ratio is substantially larger than that observed for other imaged planetary systems such as HR 8799 
(q $\sim$ $5*10^{-3}$; \citealt{fabrycky2010}) and ROXs 42B 
(q $\sim$ 0.008-0.01; \citealt{cur14}).
Rather, it is more similar to that observed for low 
mass BDs (q $\sim$ 0.01-0.9; eg. see Figure 4, \citealt{cur14} and citations there in).  
We suggest this is indicative that the system formed 
via some form of fragmentation, i.e. a binary-star-like 
formation mechanism, rather than core accretion \citep{pollack1996}.
Stone et al. \citeyear{Stone2016} reached a similar conclusion of the binary-star-like formation.

\subsection{Atmospheric Modeling}
Although VHS 1256 b occupies a similar near-IR color-magnitude space to HR 8799 bcde \citep{gau15}, its 
significantly older age than the HR 8799 system enables one to probe a different time frame in planet/brown dwarf atmospheric evolution.  VHS 1256 b and HR 8799 bcd(e?) have different spectral energy distributions at the longest wavelengths probe ($M^\prime$/4.7 $\mu m$).   In the now-standard picture of understanding the atmospheres of the youngest and lowest-mass L/T objects, thick clouds and non-equilibrium carbon chemistry both are due to the objects' low surface gravity \citep[e.g.][]{Marley2012}.  That VHS 1256 b thus far lacks evidence for non-equilibrium carbon chemistry may complicate this picture, suggesting some decoupling of gravity's two effects or that VHS 1256 b's gravity is high enough that non-equilibrium effects are less obvious than they are for, say, HR 8799 b.  

Higher signal-to-noise detections in $M^\prime$ and photometry in the 3--4 $\mu m$ range probing methane will allow us to better clarify VHS 1256 b's carbon chemistry.  Multiple lines in $J$ band resolvable at medium resolution could better clarify the companion's surface gravity.   With other, similar objects detected at a range of ages, we can better map out the atmospheric evolution of objects of a given mass as well as the diversity of objects occupying the same reddened L/T transition region where VHS 1256 b and bona fide planets like HR 8799 bcde reside.
\subsection{System Architecture}

The VHS 1256 hierarchical triple system is poised to become an important contributor to our understanding 
of VLM systems.  It represents the third hierarchical triple system comprised solely of brown dwarf-mass 
components known \citep{bou05,rad13}.  Given this projected separation (1.3 au) and associated approximate orbital period ($\sim$4.7 years) of the central binary in VHS 1256, future AO spectroscopic monitoring of the system is poised to determine dynamical masses of all components 
of the triple system, which should help constrain evolutionary models (see e.g. \citealt{dupuy2010}).  Since 
at least some brown dwarf binaries are believed to form via the disintegration of triple systems, and the 
third body in such systems are most likely also brown dwarf mass object \citep{rei15}, robustly determining 
the fundamental properties of the few known triple systems like VHS 1256 could help test the predictions of 
dynamical simulations of BD formation and evolution.
\\

We thank Sarah Schmidt for thoughtful discussions about L/T dwarfs and multiplicity and Jonathan Gagne for helpful comments on VHS 1256's possible membership in different moving groups.  
We acknowledge support from NSF-AST 1009314 and NASA's Origins of Solar Systems 
program under NNX13AK17G.  This work was performed in part under contract with 
the Jet Propulsion Laboratory (JPL) funded by NASA through the Sagan 
Fellowship Program executed by the NASA Exoplanet Science Institute. This work is also based 
on data collected at Subaru Telescope, which is operated by the National Astronomical 
Observatory of Japan.  The 
authors recognize and acknowledge the significant cultural role and 
reverence that the summit of Mauna Kea has always had within the indigenous 
Hawaiian community. We are most fortunate to have the opportunity to 
conduct observations from this mountain.

%----------------------------------------------------------- FIGURES

\clearpage

\pagebreak

\begin{table}
\begin{center}
\caption{VHS1256 Astrometry and Magnitudes}\label{tbl:properties}
\tablewidth{0pc}
\tablecolumns{3}
\begin{tabular}{crrr}
\\
\tableline\tableline
VHS 1256 Objects and Band & \textbf{System Properties} \\
\tableline

A and B separation ($L^\prime$) & 0$\farcs$1056 $\pm$ 0$\farcs$001 \\
A and B PA ($L^\prime$) & $-6^{\circ} \pm 2^{\circ}$ \\
A and B Est. Period ($L^\prime$) & 4.7 yrs \\
A apparent magnitude ($L^\prime$) & 10.50 $\pm$0.01 mag \\
B apparent magnitude ($L^\prime$) & 10.54 $\pm$0.01 mag \\
A+B and b separation ($L^\prime$) & 8."13 $\pm$ 0."04 \\
A+B and b PA ($L^\prime$) & 217.8 $\pm$ 0.3 $^{\circ}$ \\
A+B and b separation ($M^\prime$) & 8."17 $\pm$ 0."04 \\
A+B and b PA ($M^\prime$) & 217.8 $\pm$ 0.3 $^{\circ}$ \\
\tableline
\end{tabular}
\tablecomments{The measured and derived for VHS 1256 A, B, and b in $L^\prime$and $M^\prime$ using centroid positions and PSF fitting.}
\end{center}
\end{table}

\begin{deluxetable}{llcccc}
 \tiny
\tabletypesize{\tiny}
\tabletypesize{\small}
\tablecolumns{3}
\tablecaption{VHS1256 Aperture Photometry}
\tablehead{{Photometry Band}& {VHS 1256 A and B} & {VHS 1256 b}\\
&(mag) &(mag)}
\startdata
L' & 9.76 $\pm$ 0.03 & 12.99 $\pm$ 0.04 \\
M' & 9.65 $\pm$ 0.05 & 12.66 $\pm$ 0.26 \\
W1 & 9.880 $\pm$ 0.023 & 13.6 $\pm$ 0.5 \\
W2 & 9.658 $\pm$ 0.021 & 12.8 $\pm$ 0.5 \\
 \enddata
\tablecomments{L' and M' aperture photometry taken from traditional reductions described in section \ref{sec:obs} with the exception of M' photometry of VHS 1256 b which used the LOCI reduction described in section \ref{sec:obs}. The WISE data was taken from \citet{gau15}.}
\label{tbl:photometry}
\end{deluxetable}

\begin{deluxetable}{llcccc}
 \tiny
\tabletypesize{\tiny}
\tabletypesize{\small}
\tablecolumns{11}
\tablecaption{Atmosphere Model Fitting Results (Freely-Varying Radius)}
\tablehead{{}&{  Best-Fit Model}&{Models Matching Within 3-$\sigma$}\\
{Model}& {T$_{eff}$ (K), log(g), Radius ($R_{J}$)} & {T$_{eff}$ (K), log(g), Radius ($R_{J}$)}}
\startdata
Burrows/A60 &1000, 4.5, 1.34  & 1000, 4.5, 1.34 \\

Burrows/AE60 &  800, 3.8, 1.8 & 700, 3.5--4, 2.57-2.64 \\
           &               & 750, 3.5--4.3, 2.03-2.25\\
           &               & 800, 3.5--4.5, 1.74-1.95\\
           &               & 850, 3.5--4.3, 1.60-1.64\\
           &                     & 900, 3.5-3.9, 1.35, 1.42\\
Burrows/E60      & 1000, 4.0, 0.98 & -- \\

 \enddata
\tablecomments{Atmosphere modeling results shown in increments of 50 $K$.  The radii quoted assume the revised parallax from \citet{Stone2016}.  Owing to uncertainty in VHS 1256-12's parallax from that paper, the best-fit radii have an additional systematic uncertainty of $\approx$ 10\%.  Radii assuming a nominal distance of 12.7 $pc$ are systematically $\approx$ 35\% smaller.}
\label{tbl:atmosfit}
\end{deluxetable}

\begin{deluxetable}{llcccc}
 \tiny
\tabletypesize{\tiny}

\tabletypesize{\small}
\tablecolumns{11}
\tablecaption{Atmosphere Model Fitting Results (Fixed Radius)}

\tablehead{{}&{ Best-Fit Model}&{Models Matching to Within 3-$\sigma$}\\
{Model}& {T$_{eff}$ (K), log(g), Radius ($R_{J}$)} & {T$_{eff}$ (K), log(g), Radius ($R_{J}$)}}
\startdata
\textit{d = 17.1 $pc$}\\
Burrows/A60 &1000, 4.25, 1.45  & -- \\

Burrows/AE60 &  850, 3.8, 1.55 & 850, 3.5--3.9, 1.48--1.75 \\

Burrows/E60      & 700, 4.0, 1.45 & -- \\\\
\textit{d = 12.7 $pc$}\\
Burrows/A60 &900, 4.25, 1.29  & -- \\

Burrows/AE60 &  800, 4.1, 1.36 & 750, 3.5--3.8, 1.53--1.73\\
           &               & 775, 3.8--4.1, 1.35--1.53\\
           &               & 800, 4.0--4.3, 1.25--1.41\\
           &               & 850, 4.3, 1.26\\

Burrows/E60      & 700, 4.0, 1.45 & -- \\

 \enddata
\tablecomments{Atmosphere modeling results, shown in increments of 25 $K$. As before, for a given parallax assumption the best-fit radii have an additional systematic uncertainty of $\approx$ 10\%.  }

\label{tbl:atmosfitfixed}
\end{deluxetable}

\begin{table}
\begin{center}
\caption{VHS1256 Absolute Magnitudes and Masses }\label{tbl:absolute}
\tablewidth{0pc}
\tablecolumns{3}
\begin{tabular}{crrr}
\\
\tableline\tableline

Object & Measurement  & Measurement  \\
 & at 12.7 $\pm$ 1.0 pc & at 17.1 $\pm$ 2.5 pc \\
\tableline
\\
A Absolute magnitude ($L^\prime$) & 10.0 $\pm$ 0.2 mag & 9.3 $\pm$ 0.3 mag\\
B Absolute magnitude ($L^\prime$) & 10.0 $\pm$ 0.2 mag & 9.4 $\pm$ 0.3 mag\\
Age lower limit & $>$ 400 Myrs & $>$ 200 Myrs \\
A and B Mass BT-Settll$^{a}$ (L') & 58 M$_J$ & 58 M$_J$ \\
b Absolute Magnitude (L') & 12.5 $\pm$ 0.2 mag & 11.8 $\pm$ 0.3 mag \\
b Mass BT-Settl$^{a}$ (L') & 26.2 to 74 M$_J$ & 15.7 to 75 M$_J$ \\
b Bolometric Luminosity log(L/L$_{\sun}$) & -5.06 to -5.24 & -4.79 to -4.95 \\
b Mass BT-Settl$^{a}$ Bolometric & 12.6 - 20.6 M$_J$ & 10.5 - 15.7 M$_J$ \\
Limiting Absolute Magnitude at 2$\farcs$8 (L') & 15.6 & 14.9 \\
Limiting Magnitude Mass SB12$^{b}$ models (L') & $\sim$ 10 M$_J$ & $\sim$ 10 M$_J$ \\
Limiting Magnitude Mass BT-Settl$^{a}$ models (L') & 5 M$_J$ & 7.8 M$_J$ \\
\end{tabular}
\tablecomments{The measured and derived parameters for VHS 1256 A, B, and b in $L^\prime$and $M^\prime$. a) BT-Settle model from \citet{all14}. b) SB12 models from \citet{SB12}.}
\end{center}
\end{table}

\nocite{*}

\begin{thebibliography}{9}
\bibitem[Allard(2014)]{all14} Allard, F. 2014 in Exploring the Formation and Evolution of Planetary 
Systems, IAU Proc. 299, 271
\bibitem[Anthes Rich et al.(2015)]{rich2015} Anthes Rich, E., Wisniewski, J.~P., Hashimoto, J., et al.\ 2015, AAS/Division for Extreme Solar Systems Abstracts, 3, 117.10 
\bibitem[Armstrong et al.(2014)]{arm14} Armstrong, D.J., Osborn, H.P., Brown, D.J.A., Faedi, F., Gomez Maqueo Chew, Y., Martin, D.V., Pollacco, D., \& Udry, S. 2014, MNRAS, 444, 1873
\bibitem[Bailey et al.(2014)]{bai14} Bailey, V., Meshket, T., Reiter, M. et al. 2014, ApJL, 780, 4
\bibitem[Baraffe et al.(2003)]{Baraffe2003}Baraffe, I., Chabrier, G., Barman, T. S., et al., 2003, A\&A, 402, 701 
\bibitem[Bate(2009)]{bate2009} Bate, M.~R.\ 2009, \mnras, 392, 590
\bibitem[Bell et al.(2015)]{bel15} Bell, C.P.M., Mamajek, E.E., \& Naylor, T. 2015, MNRAS, 454, 593
\bibitem[Borucki et al.(2011)]{bor11} Borucki, W.J., Koch, D.G., Basri, G. et al. 2011, ApJ, 736, 19
\bibitem[Bonnefoy et al.(2016)]{Bonnefoy2016}Bonnefoy, M., Zurlo, A., Baudino, J.-L., et al., 2016, A\&A, in press
\bibitem[Bouy et al.(2005)]{bou05} Bouy, H., Martin, E.L., Brandner, W., \& Bouvier, J. 2005, AJ, 129, 511
\bibitem[Bowler et al.(2015)]{bowler2015} Bowler, B.~P., Liu, M.~C., Shkolnik, E.~L., \& Tamura, M.\ 2015, \apjs, 216, 7 
\bibitem[Brandt et al.(2014)]{bra14} Brandt, T.D., Kuzuhara, M., McElwain, M.W. et al. 2014, ApJ, 786, 1
\bibitem[Bryan et al.(2016)]{Bryan2016}Bryan, M., et al., 2016, \apj, in press
\bibitem[Burrows et al.(2001)]{Burrows2001}Burrows, A., Hubbard, W. B., Lunine, J. I., Liebert, J., 2001, Review of Modern Physics, 73, 719
\bibitem[Burrows et al.(2006)]{Burrows2006}Burrows, A., Sudarsky, D., Hubeny, I., 2006, \apj, 640, 1063
\bibitem[Chabrier et al.(2000)]{Chabrier2000} Chabrier, G., Baraffe, I., Allard, F., \& Hauschildt, P.\ 2000, \apj, 542, 464 
\bibitem[Carson et al.(2013)]{car13} Carson, J., Thalmann, C., Janson, M., et al.\ 2013, \apjl, 763, L32 
\bibitem[Currie et al.(2011)]{Currie2011}Currie, T., Burrows, A., Itoh, Y., et al., 2011, \apj, 729, 128
\bibitem[Currie et al.(2012)]{Currie2012}Currie, T., Debes, J. H., Rodigas, T. J., et al., 2012, \apj, 760, L32
\bibitem[Currie et al.(2014a)]{cur14} Currie, T., Daemgen, S., Debes, J., Lafreniere, D., Itoh, Y., Jayawardhana, R., Ratzka, T., \& Correia, S. 2014, ApJL, 780, 30
\bibitem[Currie et al.(2014b)]{cur14B} Currie, T., Burrows, A., Girard, J.~H., et al.\ 2014, \apj, 795, 133
\bibitem[Currie et al.(2015a)]{cur15}Currie, T., Cloutier, R., Grady, C., et al., 2015, \apj, 814, L27
\bibitem[Currie et al.(2015b)]{cur15b}Currie, T., Lisse, C., Kuchner, M., et al., 2015, \apj, 807, L7
\bibitem[Duchene \& Kraus(2013)]{duc13} Duchene, G. \& Kraus, A. 2013, ARA\&A, 51, 269
\bibitem[Doyle et al.(2011)]{doy11} Doyle, L.R., Carter, J.A., Fabrycky, D.C. et al. 2011, Science, 333, 1602
\bibitem[Dupuy et al.(2010)]{dupuy2010} Dupuy, T.~J., Liu, M.~C., Bowler, B.~P., et al.\ 2010, \apj, 721, 1725
\bibitem[Dupuy \& Liu(2012)]{dupuy2012} Dupuy, T.~J., \& Liu, M.~C.\ 2012, \apjs, 201, 19
\bibitem[Hayano et al.(2008)]{hay08} Hayano, Y., Takami, H., Guyon, O. et al. 2008, in Proc. SPIE, 
Adaptive Optics Systems, Eds. N. Hubin, Claire Max, \& P. Wizinowich, Vol 7015, 10
\bibitem[Hayano et al.(2010)]{hay10} Hayano, Y., Takami, H., Oya, S. et al. 2010, in Proc. SPIE, 
Vol. 7736, 0
\bibitem[Howard et al.(2012)]{how12} Howard, A.W., Marcy, G.W., Bryson, S.T. et al. 2012, ApJS, 201, 15
\bibitem[Fabrycky \& Murray-Clay(2010)]{fabrycky2010} Fabrycky, D.~C., \& Murray-Clay, R.~A.\ 2010, \apj, 710, 1408
\bibitem[Faherty et al.(2016)]{Faherty2016} Faherty, J.~K., Riedel, A.~R., Cruz, K.~L., et al.\ 2016, arXiv:1605.07927
\bibitem[Gagne et al.(2014)]{gag14} Gagne, J., Lafreniere, D., Doyon, r., Malo, L., \& Artigau, E. 2014, ApJ, 783, 121
\bibitem[Gauza et al.(2015)]{gau15} Gauza, B., Bejar, V.J.S., Perez-Garrido, A. et al. 2015, ApJ, 804, 96
\bibitem[Galicher et al.(2011)]{Galicher2011}Galicher, R., Marois, C., Macintosh, B., et al., 2011, \apj, 739, L41
\bibitem[Hinz et al.(2010)]{hin10} Hinz, P., Rodigas, T.J., Kenworthy, M.A. et al. 2010, ApJ, 716, 417
\bibitem[Kalas et al.(2008)]{kal08} Kalas, P.G., Graham, J.R., Chiang, E. et al. 2008, Science, 322, 1345
\bibitem[Knapp et al.(2004)]{knapp2004} Knapp, G.~R., Leggett, S.~K., Fan, X., et al.\ 2004, \aj, 127, 3553 
\bibitem[Kobayashi et al.(2000)]{kob00} Kobayashi, H., Kawaguchi, N., Fujisawa, K. et al. 2000, 
in Proc. SPIE, Optical and IR Telescope Instrumentation and Detectors, eds M. Iye \& A. F. Moorwood, Vol. 4008, 1056
\bibitem[Kuzuhara et al.(2013)]{kuz13} Kuzuhara, M., Tamura, M., Kudo. T. et al. 2013, ApJ, 774, 11
\bibitem[Lafreniere et al.(2007)]{Lafreniere2007a}Lafreniere, D., Marois, C., Duyon, R., et al. 2007, \apj, 660, 770
\bibitem[Lafreni{\`e}re et al.(2008)]{Laf2008} Lafreni{\`e}re, D., Jayawardhana, R., \& van Kerkwijk, M.~H.\ 2008, \apjl, 689, L153 
\bibitem[Lagrange et al.(2009)]{lagrange2009} Lagrange, A.-M., Gratadour, D., Chauvin, G., et al.\ 2009, \aap, 493, L21 
\bibitem[Lagrange et al.(2010)]{lag10} Lagrange, A.-M., Bonnefoy, M., Chauvin, G., Apai, D., Ehrenreich, D., Boccaletti, A., Gratadour, D., Rouan, D., Mouillet, D., Lacour, S., \& Kasper, M. 2010, Science, 329, 57
\bibitem[Laughlin, Bodenheimer, \& Adams(2004)]{lau04} Laughlin, G., Bodenheimer, P., \& Adams, F.C. 2004, ApJL, 612, 73
\bibitem[Leggett et al.(2002)]{Legget2002} Leggett, S.~K., Golimowski, D.~A., Fan, X., et al.\ 2002, \apj, 564, 452
\bibitem[Leggett et al.(2010)]{Leggett2010} Leggett, S.~K., Burningham, B., Saumon, D., et al., 2010, ApJ, 710, 1627
\bibitem[Liu et al.(2013)]{Liu2013}Liu, M.C., Magnier, E. A>, Neacon, N. R., et al., 2013, \apj, 777, L20
\bibitem[Low \& Lynden-Bell(1976)]{Low1976} Low, C., \& Lynden-Bell, D.\ 1976, \mnras, 176, 367
\bibitem[Macintosh et al.(2015)]{mac15} Macintosh, B., Graham, J.R., Barman, T. et al. 2015, Science, 350, 64
\bibitem[Madhusudhan et al.(2011)]{Madhusudhan2011}Madhusudhan, N., Burrows, A., Currie, T., 2011, \apj, 737, 34
\bibitem[Malo et al.(2013)]{mal13} Malo, L., Doyon, R., Lafreniere, D., et al. 2013, ApJ, 762, 88
\bibitem[Marley et al.(2012)]{Marley2012}Marley, M., Saumon, D., Cushing, M., et al., 2012, \apj, 754, 135
\bibitem[Marois et al.(2010)]{mar10} Marois, C., Zuckerman, B., Konopacky, Q.M., Macintosh, B., \& Barman, T. 2010, Nature, 468, 1080
\bibitem[Naud et al.(2014)]{nau14} Naud, M.-E., Artigau, E., Malo, L. et al. 2014, ApJ, 787, 5
\bibitem[Quanz et al.(2013)]{quanz13}Quanz, S., Meyer, M. R., Kenworthy, M., et al., 2013, ApJ, 766, L1
\bibitem[Pollack et al.(1996)]{pollack1996} Pollack, J.~B., Hubickyj, O., Bodenheimer, P., et al.\ 1996, Icarus, 124, 62 
\bibitem[Radigan et al.(2013)]{rad13} Radigan, J., Jayawardhana, R., Lafreniere, D. et al. 2013, ApJ, 778, 36
\bibitem[Raghavan et al.(2010)]{rag10} Raghavan, D., McAlister, H.A., Henry, T.J. et al. 2010, ApJS, 190, 1
\bibitem[Ratzka, Kohler, \& Leinert (2005)]{rat05} Ratzka, T., Kohler, R., \& Leinert, C. 2005, A\&A, 437, 611
\bibitem[Reipurth \& Mikkola(2015)]{rei15} Reipurth, B. \& Mikkola, S. 2015, AJ, 149, 145
\bibitem[Simon et al.(1995)]{sim95} Simon, M., Ghez, A.M., Leinert, C. et al. 1995, ApJ, 443, 625
\bibitem[Skemer et al.(2012)]{ske12} Skemer, A., Hinz, P.M., Esposito, S. et al. 2012, ApJ, 753, 14
\bibitem[Spiegel \& Burrows(2012)]{SB12} Spiegel, D.~S., \& Burrows, A.\ 2012, \apj, 745, 174
\bibitem[Stone et al.(2016)]{Stone2016} Stone, J., et al., 2016, \apj\L, 818, 12
\bibitem[Thalmann et al.(2014)]{tha14} Thalmann, C., Desidera, S., Bonavita, M. et al. 2014, A\&A, 572, 91
\bibitem[Todorov et al.(2010)]{trodorv2010} Todorov, K., Luhman, K.~L., \& McLeod, K.~K.\ 2010, \apjl, 714, L84
\bibitem[van der Bliek, Manfroid, \& Bouchet(1996)]{van96} van der Bliek, N.S., Manfroid, J., \& Bouchet, P. 1996, A\&AS, 119, 547
\bibitem[Winn \& Fabrycky(2015)]{win15} Winn, J.N. \& Fabrycky, D.C. 2015, ARAA, 53, 409


\end{thebibliography}
\end{document}